# Structural properties of Fe/Cu magnetic multilayers: a Monte Carlo approach

Jules Berlin Nde Kengne[1,2], Bernard Fongang[2,3], and Serge Zekeng[1,3]

[1] Laboratory of Material Science, Department of Physics, Faculty of Science, University of Yaoundé I, P.O. Box 812 Yaoundé, Centre, Cameroon.

[2] Department of Biochemistry and Molecular Biology, University of Texas Medical Branch at Galveston, Galveston, Texas, USA.

[3] Correspondence should be addressed to S.Z. (sergezekeng@yahoo.fr) or B.F. (befongan@utmb.edu)

**Abstract**

Using atomistic Monte Carlo simulations, we investigated the impact of the interface on the structural properties of iron and copper (Fe/Cu) magnetic multilayers grown by Voronoï diagram. Interest in magnetic multilayers has recently emerged as they are shown to be promising candidates for magnetic storage media, magneto-resistive sensors, and personalized medical treatment. As these artificial materials show large differences in properties compared to conventional ones, many experimental and theoretical works have been dedicated on shedding light on these differences and tremendous results have emerged. However, little is known how the interfaces influence the structure of the layers around them. By a numerical approach, we show that the structure of each layer depends on its thickness and the interface morphology. The Fe and Cu layers can adopt either the body-centered-cubic (bcc) or face-centered-cubic (fcc) structure, while the interface can assume amorphous, bcc, fcc, or a mixture of bcc and fcc structures depending on the layer thicknesses. These results are in good agreement with the experiments. They could be helpful in understanding effects such as giant magneto-resistance from the structural viewpoint.

**Introduction**

Metallic multilayer systems have gain much interest as they show very different properties from bulk materials.[1,2] These artificial materials not only advance our understanding of the fundamental physics of nanomagnetism but also have a wide range of technological applications such as magnetic recording media, magnetic storage, and magneto-resistive sensors.[1-4] Despite many experimental and theoretical progresses,[5-9] our understanding of those differences is still incomplete.

Many factors can contribute to the differences between multilayer systems and conventional bulk materials.[6,10] One likely factor is the exchange coupling between thin films of a ferromagnetic metal separated by non/anti-ferromagnetic metal. The exchange coupling could explain the phenomena called giant magneto-resistance (GMR).[1,11,12] GMR was first observed in Fe/Cr multilayer systems: when varying the thickness of the Cr layer between two Fe layers, the magneto-resistance changed considerably.[13-15] This phenomenon was later observed in other magnetic multilayers such as Fe/Cu.[6,7,16,17] However, the spacer thickness to which the oscillating coupling or the maximum of GMR is observed remains to be addressed. Moreover, the increasing percentage of atoms at the interface with the decreasing layer thicknesses can make structural properties of metallic multilayer systems to deviate significantly from those of their bulk counterparts.[18,19] In addition to the surface and interface effects, the electronic transport through multilayer structures can also affect the properties of magnetic multilayers.

Although many techniques have been developed for the preparation and characterization of magnetic multilayers, there are subtle differences among those techniques that can sometimes lead to misinterpretations of results.[20-23] For example, various deposition rates and growth temperatures can produce unexpected structural changes in the multilayer system.[24] Furthermore, soluble materials may experience unwanted interdiffusion at the interface. (To reduce the interdiffusion, multilayer systems made of Fe and Cu are often chosen for their mutual insolubility.[25])

Besides the experimental approaches, numerical methods provide an alternative to study these systems.[9, 26] However, it remains challenging to accurately model the structure of the system, mainly at the surface and interface, which critically influences the structure of the rest of the system and the overall magnetic properties.[27, 28] It is then desirable to have a numerical method of growing multilayer systems that better mimic experimental samples.

In this paper, we present a numerical study on the structural properties of metallic multilayers with variable layer thicknesses. We use Fe/Cu multilayer systems as a model and investigate the structural changes using atomistic Monte Carlo simulations, which can describe the properties of magnetic multilayers successfully.[29, 30] We use the Voronoï diagram method[31] to construct the numerical sample in order to make better comparison with the experiment.

**Methodologies**

*Construction of multilayer systems via Voronoï diagram*

We built multilayer systems based on Voronoï diagram.[19, 31] Dirichlet and Voronoï first introduced the Voronoï diagram in geometry,[32] and it was later related to Delaunay triangulation. The construction of a Voronoï diagram consists of partitioning a plane into regions based on distance to points in a specific subset of the plane. This set of points (called crystallographic cell) is specified beforehand, and for each cell, there is a corresponding region consisting of all points closer to that cell than to any other. These regions are called Voronoï cells. The Voronoï diagram of a set of points is dual to its Delaunay triangulation. The general idea here is that given a set $S$ of $n$ point sites in the d-space, Voronoï diagram of $S$, $V_k(S)$, partitions the space into regions such that each point within a fixed region has the same $k$ closest sites.

We grew our multilayer systems according to the principle that each layer is characterized by his nucleation center of coordinate ($X_i, Y_i, Z_i$, where $i$ = 1 to $n$, $n$ stands for the number of layers) in the direct orthogonal base (O, I, J, K, where O stands for the origin, and I, J, or K standing for the unit coordinate for $X, Y,$ or $Z$ respectively). For each center, Euler angles are chosen in order to orient the relative crystallographic axes of the layers. This arrangement aims to produce the initial step for each layer. The next step is to add atoms to the system, to make the layers grow in all directions until the structure reaches a certain limit fixed by the Voronoï cell conditions. The detailed procedure used in this study can be found elsewhere.[19] We further avoided finite size and surface effects by first choosing a simulation box large enough to contain several layers and then performed simulations on a truncated box of much smaller size.

*Energy function and Monte Carlo simulations*

To mimic realistic samples, we used Modified Embedded Atom Method (MEAM) for the energy function of the system. While the current energy function can partially describe certain physical properties, it still suffers from certain accuracy problems. For example, Byeong et al.[33, 34] showed that the results from MEAM for many bcc metals were in contradiction with the experimental ones;[35, 36] they also showed that for a bcc structure, the energy of the (111) surface was smaller than that of the (100) surface.

The presence of multiple elements at the interface makes the problem even more complex for studying multilayer systems. Therefore, to overcome these difficulties, we used the variant of MEAM developed by Baskes et al.,[37, 38] for pure and alloy elements. This potential showed good agreement with the experiments in describing physical properties and surface energies.[38] Compared to EAM, MEAM has the advantage of being able to reproduce a wide range of structures including fcc, bcc, hcp, etc., for different elements, using one common functional form. The following relation gives the form of the total energy used in this study:

$$E = \sum_i [F_i(\rho_i) + \frac{1}{2}\sum_{j(\neq i)} S_{ij}\Phi_{ij}(R_{ij})] \quad (1)$$

where $F_i$ is the embedding function, $\rho_i$ the background electron density at site $i$, $S_{ij}$ the screen factor, and $\Phi_{ij}$ the pair interaction between atoms $i$ and $j$ separated by a distance $R_{ij}$. More details about the potential can be found elsewhere.[38]

To test the accuracy of the potential, we applied it to an alloy system composed of 50 % of Fe and 50 % of Cu atoms (data not shown). The final structure after Monte Carlo simulated annealing scheme showed that the bonds among atoms are conserved inside and outside the simulation box. Moreover, the fact that Fe and Cu atoms do not form agglomerates of each type or the presence of atoms inside the simulation box after the simulated annealing Monte Carlo scheme justifies the choice of our interatomic potential.

We further minimized the total energy of the system given in equation (1) using Monte Carlo simulations with the Metropolis algorithm. Importance sampling Monte Carlo known as Metropolis Monte Carlo algorithm is based on the algorithm proposed by Metropolis et al.[39] We fixed the initial system at a high temperature (> 800K) to allow the movement of each atom and we gradually decreased the temperature (as described in[31]) until we reached the equilibrium, where the energy no longer changes (final temperature lower than 1K). For each step of temperature, we performed the following steps:

1- Calculate the energy of the system at a high temperature (initial state).
2- Randomly choose and displace an atom in any of the three directions Δx, Δy, or Δz.
3- Calculate the energy change ΔE due to the displacement and decide whether to accept the move based on Metropolis criterion:
   If $\Delta E < 0$, then accept the new configuration.
   If $\Delta E > 0$, then calculate $p = \exp(-\frac{\Delta E}{K_b T})$ (where $K_b$ represents the Boltzmann constant and T the absolute temperature) and compare it with a random number $u$ between 0 and 1. If p > u, accept the new configuration, otherwise keep the old one and restart at step 2.
4- Redo steps 1-3 for each block of temperature.

We adapted initial conditions from;[19] in addition, for complexity reason and in order to keep the structure close to that of bulk material far from the interface, we made the following modification: the choice of the atom to be subjected to the Metropolis algorithm depends on its distance to the closest interface. The following relation gives the choice's probability

$$\omega = \exp(-\alpha x) \quad (2)$$

where $\alpha$ is an adjustable parameter comprised between 0.5 and 2 $\text{Å}^{-1}$ as shown in a previous work,[31] and $x$ represents the distance from the considered atom to the nearest interface. This probability allows one to avoid periodic boundaries conditions and to anneal the atomic structure only close to the interfaces, as atoms located far from the interface have a low but non-zero probability to be chosen for the move. Interestingly, we have observed the maximum change of energy to be located only at the interface; the change far from the interface was almost negligible after we subjected the system to the simulated annealing Monte Carlo scheme (Fig. 1c, 1d).

*Structural characterization*

We further used a radial distribution function (RDF) to reproduce the change observed on the structures of layers and interfaces. A RDF $g(r)$ is the probability of finding an atom at the distance r of another identical atom chosen as reference. It well-describes the long-range order. The following relations give the distances ($d_{Fe1}$, $d_{Fe2}$, $d_{Cu1}$, and $d_{Cu2}$), between first and second nearest neighbors for pure Fe and Cu structures respectively: $d_{Fe1} = a\sqrt{3}/2$, $d_{Fe2} = a$, $d_{Cu1} = a\sqrt{2}/2$, and $d_{Cu2} = a$, where a stands for lattice parameter of Fe (~ 2.856 Å) or Cu (~ 3.597 Å). The obtained pure structures will be used in the following section as references.

**Results and discussion**

In general, we numerically grew multilayers with variable thicknesses, set the adjustable parameter $\alpha$ to 2 Å$^{-1}$, and performed 5×10$^5$ Monte Carlo steps (MCSs) at each temperature. All the results were obtained after convergence of the Metropolis annealing procedure (no considerable change was observed in energy after a decrease of temperature).

We first mimicked the bulk structures of pure elements contained in our sample. Figure 1 shows the RDF obtained for Fe and Cu layers before (a, b) and after (c, d) simulated annealing. For each layer, one can see that the positions of peaks match the above-mentioned theoretical values (first and second nearest neighbor) for conventional elements of Fe and Cu structures. These results show the suitability of our method to mimic pure and experimental samples.

While many experimental studies in literature have been dedicated to bilayer systems,[3, 8] we wondered whether the number of interfaces could have an impact on the structure of the system. We therefore first studied systems composed of two layers and one interface. In order to gain insight into the role played by the interface on the structure of bilayers Fe/Cu, we proceeded to the analysis of Fe$_5$/Cu$_{15}$, Fe$_{15}$/Cu$_5$, and Fe$_{35}$/Cu$_{35}$ (the subscripts denote thicknesses given in Å) bilayer systems. Figure 2 shows the RDF of the structures obtained after the Monte Carlo simulated annealing scheme. One can see from Figure 2a (case of Fe$_5$/Cu$_{15}$ system) that the system conserves the structure of the thickest layer, here fcc Cu. A structure similar to that of fcc Cu is also observed at the interface (Figure 2b), while the structure of Fe layer remains close to that of fcc Cu for short range interactions (Figure 2a) and amorphous far from the interface Cu/Fe. The absence of long-range order can be explained by the fact that almost all Fe atoms are located at the interface because of the ultrathin thickness of that layer, such a way that Fe atoms are embedded in the dominant element (here Cu), as interface thickness of Fe/Cu was shown to be around 4 Å.[9] Moreover, a similar behavior is observed when switching the thicknesses of Fe and that of Cu (Fe$_{15}$/Cu$_5$ system). Figure 2c shows the well-defined bcc structure of the Fe layer; interface structure (Figure 2c) is similar to that of bcc Fe and the structure of Cu is amorphous far from the interface. We concluded that for bilayer systems, the thickest layer imposes its structure to the system when the smallest layer has a thickness less than 15 Å.

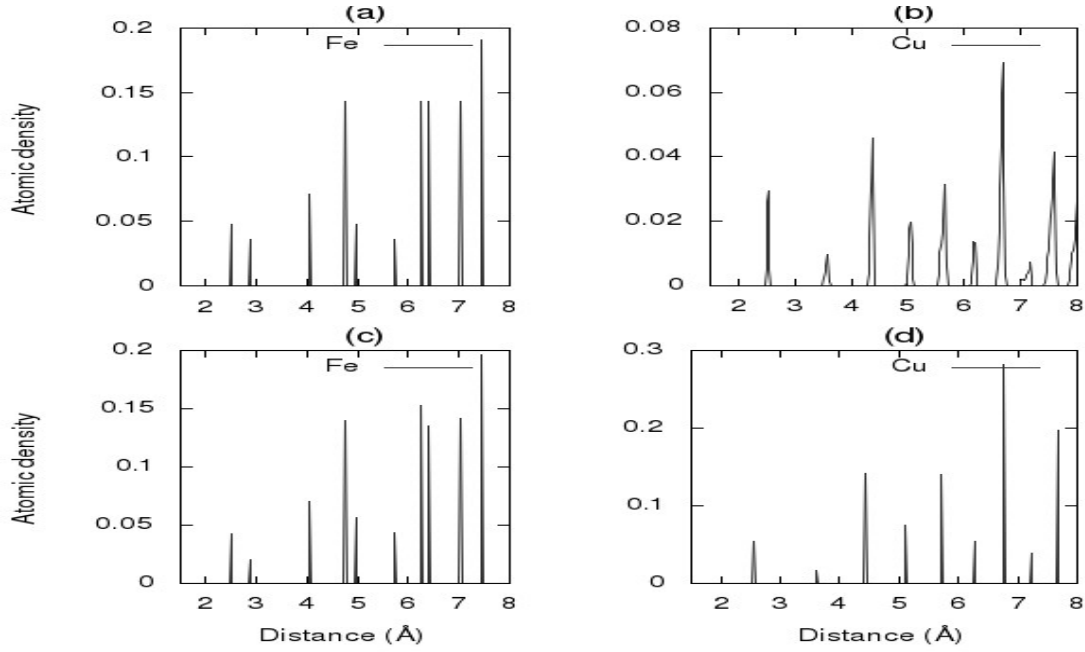

Figure 1. Radial distribution functions (RDF) of pure elements Fe (a) and Cu (b) before and Fe (c) and Cu (d) after simulated annealing. The similarities between both structures before and after Monte Carlo simulated annealing justifies the fact that we neglected the relaxation of the whole system, thus, limiting computational time.

Ultrathin layers first grow according to the structure of thickest layer (first atomic planes) and smoothly relax towards its bulk structure when increasing its thickness. We confirmed this observation when increasing the thicknesses of Fe and Cu to 35 Å (Figure 2d for multilayer $Fe_{35}/Cu_{35}$). One can see the well-pronounced fcc and bcc structures of Cu and Fe layers respectively, while the interface tends to adopt an intermediary structure. We can justify the presence of the intermediary structure at the interface instead of amorphous interface by the fact that the atomic proportion at the interface decreases with the increase in layer thicknesses as previously reported,[18, 19] and atoms tend to be surrounded by the particles of same type. Albini et al.[8] and Jian-Tao et al.[40] previously observed similar results.

We next went ahead and studied the system composed of multiple interfaces. We know from the literature that giant magneto-resistance (GMR) is commonly obtained from systems composed of three layers[11-13]. To study the impact of the second interface on the structure of the system and to accurately estimate the thicknesses of the spacer layer to which the maximum GMR could be observed, we carried out our investigations on trilayer systems. In order to further gain insight into the interfaces morphology and its impact on the structure of the system, we analyzed a series of multilayer systems, which showed a strong dependence of the structure at the interface on the thickness of layers and vice versa.

For instance, Figure 3 presents the RDF obtained from the system $Fe_{21}/Cu/Fe_{21}$, with variable spacer thickness: $t_{Cu}$ = 6, 13, 24, and 40 Å.

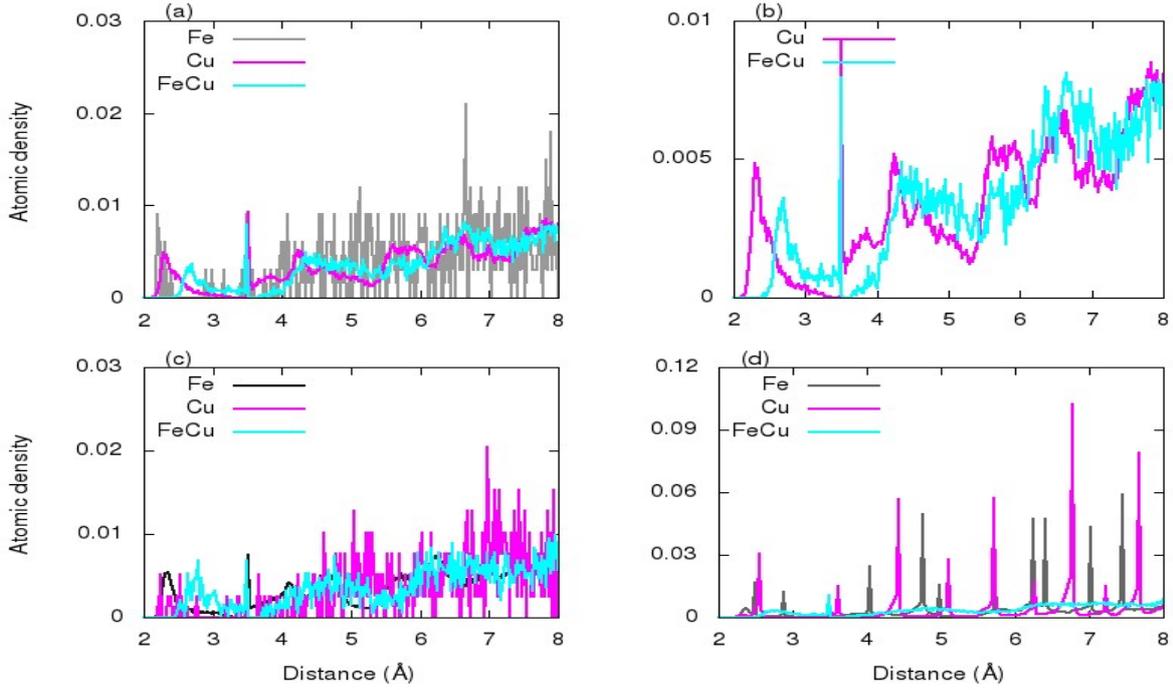

Figure 2. RDF showing the structures of bilayers Fe/Cu systems for variable thickness of Fe ($t_{Fe}$) and Cu ($t_{Cu}$) layers. a) With $t_{Fe}$ = 5 Å and $t_{Cu}$ = 15 Å, there is a structural similarity between the Cu layer and the interface, as confirmed in (b) (the structure of Fe layer is close to fcc for the first planes and amorphous beyond); c) With $t_{Fe}$ = 15 Å and $t_{Cu}$ = 5 Å, the Cu layer is amorphous, and both Fe layer and interface adopt a bcc-like structure; d) With $t_{Fe}$ = 35 Å and $t_{Cu}$ = 35 Å, there are well pronounced structures of Fe (bcc) and Cu (fcc) layers and an intermediary structure at the interface

One can see that for the ultrathin spacer (Figure 3a, system $Fe_{21}/Cu_6/Fe_{21}$), the structure of Fe layer remains bcc, while that of Cu layer tends to copy the metastable bcc structure for the first atomic planes and adopts an amorphous structure beyond. We observed the presence of Fe atoms even at the core of Cu layer (data not shown), confirming the fact that Cu layer is not large enough to overcome the interface effect (i.e. its structure is more dominated by the interface structure, here bcc). Moreover, the lack of proper Cu ordered structure even at the core of Cu layer could also be related to the presence of two interfaces, as $t_{Cu}$ is not large enough to keep a single domain containing only Cu atoms. Compared to the results on Figure 2c, the metastable bcc Cu structure is well pronounced. This observed structure is probably the consequence of the second interface. While it is known that Fe and Cu are immiscible materials, it has been shown that Fe and Cu can form a mixture of few planes at the interface,[41] supporting a non-pronounced structure of bulk Cu for this system. The interface structure remains similar to that of bcc structure, as one can see from Figure 4a. The peaks' positions are similar for both structures of Fe and interface of Fe/Cu.

Given that the above-mentioned structures of Cu and interface are still dominated by thickest layer (here Fe), we increased the spacer thickness and we observed the change on both structures (Figure 3b, system $Fe_{21}/Cu_{13}/Fe_{21}$). The structure of Cu tends to relax towards its fcc bulk structure while the peaks observed on the interface structure become broader. The observed structure at the interface (Figure 4b) can be explained by the fact that, increasing the spacer thickness leads to the decrease in the atomic proportion at the interface,[18, 19] hence a tendency to form an interface mixture as atoms at the interface form bonds quasi identical to those of the conventional material. We further noted that the order at the interface increases with the increase in $t_{Cu}$, this can also be explained by the fact that the density of Cu atoms at the interface reduces when $t_{Cu}$ increases,[19] therefore Cu atoms tend to form fcc Cu-like structure and a mixture of bcc and fcc structures tend to be formed at the interface. The structure of Cu layer is well pronounced with the increase in $t_{Cu}$ (Figure 3 c, d for the systems

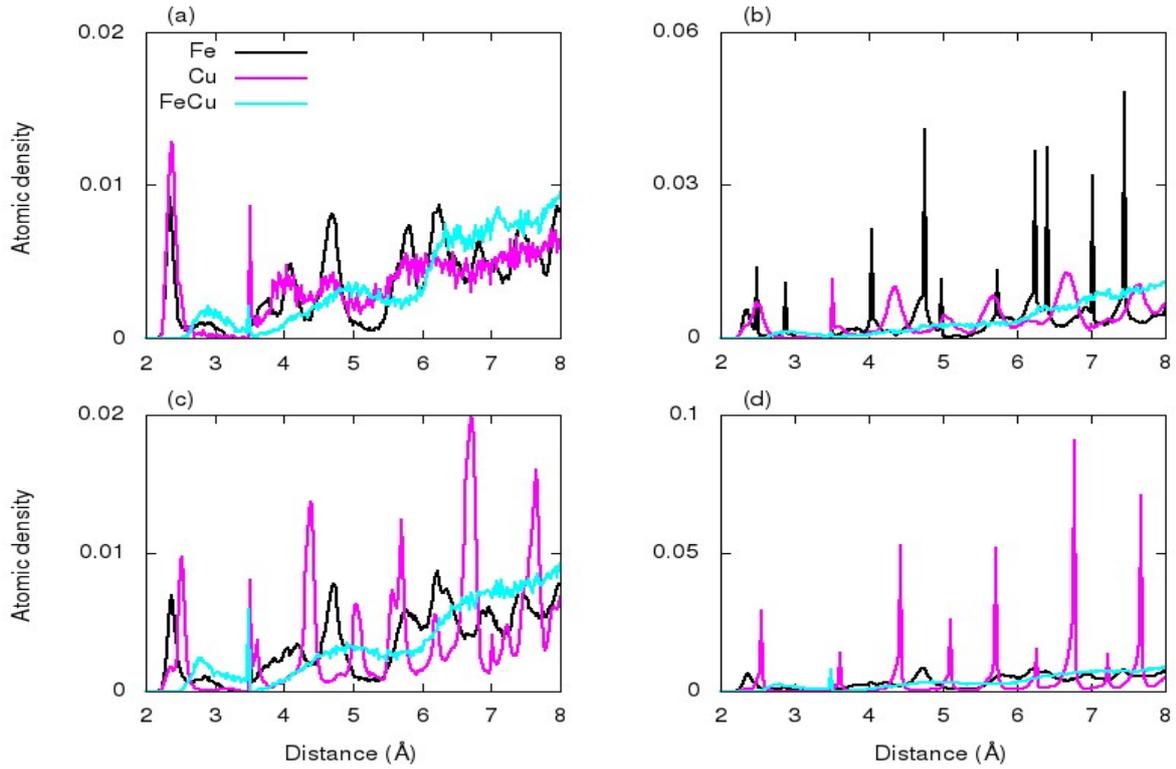

Figure 3. Multilayer systems Fe/Cu/Fe with fixed thickness of Fe layer $t_{Fe}$ = 21 Å and a variable thickness of the Cu layer $t_{Cu}$: a) with $t_{Cu}$ = 6 Å, the system adopts a bcc structure similar to that of Fe; b) with $t_{Cu}$ = 13 Å, the Cu layer undergoes a transition from metastable bcc to a structure similar to fcc of the bulk Cu, and the interface is largely amorphous; c) with $t_{Cu}$ = 24 Å, the bcc structure of the Fe layer coexists with the fcc structure of the Cu layer at the interface; d) shows for $t_{Cu}$ = 40 Å the effect of the interface morphology and spacer thickness on the structure of Fe layer

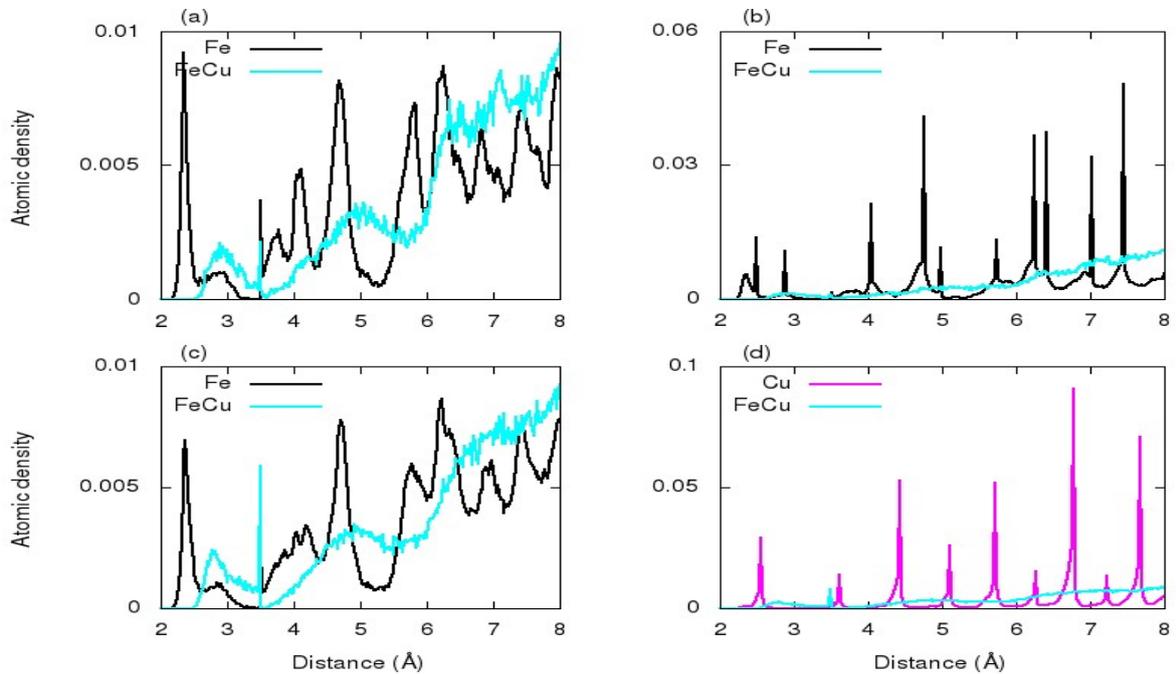

Figure 4. Multilayer $Fe_{21}/Cu/Fe_{21}$ systems, with the thickness of the Cu layer varying from $t_{Cu}$ = 13 Å to $t_{Cu}$ = 40 Å. (a) Conserved bcc structure is observed at the interface; (b) critical point denoting the transition with an amorphous structure at the interface ($t_{Cu}$ = 13 Å); (c) well-pronounced intermediary bcc and fcc interface for $t_{Cu}$=24 Å; (d) shows the conserved structure of Fe and Cu with intermediary interface structure $t_{Cu}$ = 40 Å

Fe$_{21}$/Cu$_{24}$/Fe$_{21}$ and Fe$_{21}$/Cu$_{40}$/Fe$_{21}$ respectively) confirming the above observations. It can be seen on Figure 3c that the observed Cu structure is similar to that of pure Cu shown in Figure 1a and b. This indicates that the spacer layer is large enough to reproduce the structure of bulk material; no change was observed with a further increase in spacer thickness (Figure 3d). Surprisingly, a considerable change was observed on the peaks' height of Fe structure when increasing the spacer thickness beyond a certain value, here 24 Å (Figure 3c, d). This observed behavior on the Fe layer structure can be explained by the fact that an increase in spacer thickness leads to the growth of first atomic planes (close to the interface Cu/Fe) according to the fcc structure and a smooth relaxation towards the bulk structure when approaching the core material. These observations will be further explained below. The structure at the interface remains intermediary between bcc and fcc as shown in Figure 4 c, d. These observations are in good agreement with the experiments.[5, 8]

To further investigate the correlation between interface morphology and the layers in presence, we studied systems with variable Fe thickness. Figure 5 shows the RDF for multilayer systems Fe/Cu$_{25}$/Fe, with Fe layer's thickness being varied: $t_{Fe}$ = 7, 18, 25, and 34 Å. At the fixed $t_{Cu}$, one can see that the structure of Fe depends on its thickness and the interface morphology. For example, when $t_{Fe}$ = 7 Å, the structure of Fe layer tends to copy the fcc structure as it was shown experimentally[3] (Figure 5a). The structure of the spacer layer, Cu, remains fcc as well as that at the interface. When increasing $t_{Fe}$ from 7 to 34 Å (Figure 5 b-d), a smooth transition is observed on the Fe structure from the less pronounced (here fcc-like structure) to the bulk bcc Fe structure (Figure 5d) with a critical (or transitional) point observed around $t_{Fe}$ = 18 Å (Figure 5c). It is noted that the structure of Cu layer remains fcc (Figure 5 a-d) independently of Fe thickness while the interface tends to mimic the fcc structure for ultrathin Fe layers (Figure 5 a, b), and relaxes towards an intermediary structure for thicker Fe layers (Figure 5d). An amorphous interface structure is observed at the transition, $t_{Fe}$ = 18 Å (Figure 5c). These results are in accordance with those previously obtained from experiments.[6]

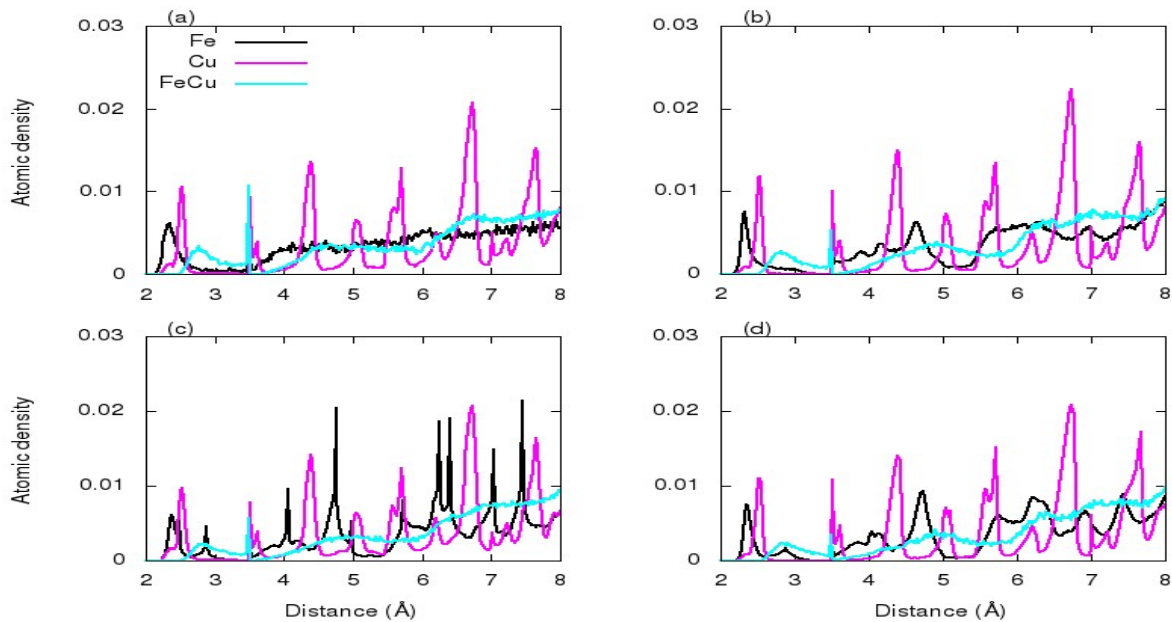

Figure 5. Multilayer systems Fe/Cu/Fe with fixed thickness of the Cu layer $t_{Cu}$ = 25 Å and variable thickness of the Fe layers, $t_{Fe}$: a) with $t_{Fe}$ = 7 Å, the system exhibits the fcc structure; b) with $t_{Fe}$ = 18 Å, the Fe layers shift towards the bulk bcc structure; c) with $t_{Fe}$ = 25 Å, the bcc structure of Fe layer is well pronounced; d) $t_{Fe}$ = 34 Å shows the well pronounced intermediary interface structure and confirms the bcc structure of the Fe layer. Cu layer structure remains fcc independently of the Fe thickness

Interface morphology plays a key role in determining the properties of magnetic multilayers. Spacer layer thickness ($t_{Cu}$) has been shown to control the nature of the interlayer exchange coupling between magnetic Fe layers during the study of the GMR effects[5-7, 13]. Depending on the spacer thickness, ferromagnetic or antiferromagnetic interlayer exchange coupling can be observed. For example, in case of ultrathin spacer layer, the loss of antiferromagnetic coupling between adjacent magnetic layers can occur due to imperfections in the spacer layer (Figure 3a) while in the case of thicker spacer layers an important reduction of the magnetic scattering can be observed due to the low concentration of Fe atoms (Figure 3d). However, the structural nature of the Cu layer and the thickness to which either of the exchange coupling is observed remain subjects of discussion. For example Figure 3b shows the structure of Cu layer similar to that of bcc Fe for the ultrathin layers, as previously observed by.[5] This implies that the direct coupling exchange between the two adjacent magnetic layers is likely to be ferromagnetic, explaining the decrease in magneto-resistance below a certain critical spacer thickness as shown in.[5] The observed metastable bcc structure of Cu can be explained by the increase in atomic proportion at the interface with the layer's thickness, and the presence at the interface of almost all Cu atoms, thus forming matrix with dominant Fe atoms. This assumption is confirmed by the structure in Figure 2c where the interface is less dominated by Fe atoms. Even though Cu layer has almost the same atomic proportion as the system represented in Figure 3a, it tends to form a Fe-Cu alloy with an amorphous structure instead of metastable bcc structure. Similarity in atomic proportion between Fe and Cu at the interface might explain this observation. The smooth transition from metastable bcc to fcc bulk Cu as observed on the graphs in Figure 3 could be at the origin of the observed expansion of the magneto-resistance with the increase of Cu thickness, for reference see.[42, 43] This transition could be helpful in explaining the observed antiferromagnetic interlayer exchange coupling for Fe/Cu/Fe multilayers.

**Conclusion**

We have investigated in this study the structural properties of metallic multilayers under the influence of the interface using atomistic Monte Carlo simulations. We have shown that the structure of the layers depends on their thickness and the structure of the interface. Moreover, the structure of the interface can adopt that of the thicker layer, and switch to an intermediary structure of fcc and bcc, as the thickness of the thinner layer increases. These results are in good agreement with the experiments.